\documentclass[%
 reprint,
]{revtex4-1}

\usepackage{graphicx}
\usepackage{dcolumn}
\usepackage{bm}
\usepackage{float}

\begin{document}

\preprint{APS/123-QED}

\title{Characterization of Thermal Neutron Beam Monitors}

\author{F. Issa}
 
\email{fatima.issa@esss.se}
\author{A. Khaplanov}%
\author{R. Hall-Wilton}%
\altaffiliation[Also at ]{Mid-Sweden University, SE-851 70 Sundsvall, Sweden}
 
\affiliation{%
 European Spallation Source ESS ERIC, P.O Box 176, SE-221 00 Lund, Sweden
}%

\author{I. Llamas}
\author{M. Dalseth Riktor}
\author{S. R. Brattheim}
\affiliation{%
 Institute for Energy Technology, P.O. Box 40 Kjeller, Norway}%

\author{H. Perrey}
\affiliation{%
Lund University, P.O Box 118, SE-221 00 Lund, Sweden}%
 \affiliation{%
 European Spallation Source ESS ERIC, P.O Box 176, SE-221 00 Lund, Sweden
}%
\date{\today}

\begin{abstract}
    Neutron beam monitors with high efficiency, low gamma sensitivity, high time and space resolution are required in  neutron beam experiments  to continuously diagnose the delivered beam. In this work, commercially available neutron beam monitors  have been characterized using the R2D2 beamline at IFE (Norway) and using a Be-based neutron source. For the gamma sensitivity measurements different gamma sources have been used.  The evaluation of the monitors includes, the study of their efficiency, attenuation, scattering and  sensitivity to gamma. In this work we report the results of this characterization. 

\end{abstract}

\pacs{Valid PACS appear here}
\maketitle

\section{Introduction}

Neutron scattering is used in modern science to understand material properties on the atomic scale. It has led to advances in many areas of science, from clean energy  to nanotechnology, materials engineering and fundamental physics. High-intensity beams of neutrons  are needed to perform neutron scattering science. These neutron beams can be produced either by fission in a nuclear reactor, similar to that at the Institut Laue-Langevin (ILL) ~\cite{ILLweb}, or by spallation sources such as at ISIS ~\cite{ISISweb}, SNS ~\cite{SNSweb}, J-PARC ~\cite{JPARCweb}. The  European Spallation Source (ESS) ~\citep{ESSweb,ESSTDR} which is under construction in Sweden, will be the world’s leading neutron source for the studies of materials. A unique feature of the ESS neutron production will be the long proton accelerator pulse (2.86 ms) with a repetition rate of 14 Hz ~\cite{Hall-Wilton2014}.

 At a  neutron spallation source, the neutron intensity fluctuates with time depending on the characteristics of the proton beam to the neutron target. This requires a continuous neutron beam monitoring with high precision to ensure the correct operation of the neutron instruments. More specifically a continuous monitoring  of neutrons in the sub-pulse structure  is required for  instruments with time-of-flight design.

 Neutron beam monitors are detectors with sufficiently low efficiency (10$^{-5}$ -10$^{-6}$) so that low percentage of the incoming  beam is absorbed or scattered. They are used to ensure that the neutron flux, beam distribution,  and pulse timing correspond to those expected from the design of the instrument. In addition, they are used to know the neutron flux at the sample in order to correctly interpret the scattering data. 

Given the diversity of the neutron instruments a variety of neutron beam monitors will be required. Multi-wire proportional counters (MWPC) have been used in many facilities as neutron beam monitors for instruments with moderate flux and low count rate requirement. They can be filled with either $^{3}$He at low pressure or nitrogen as the sensing gas. 
 For cases where a high-count rate is needed, such as for profiling on pulse-by-pulse basis, monitors based on gas electron multipliers (GEM) are more appropriate. Both types of monitors can be equipped with position sensitive readout in either 1 or 2 dimensions. Scintillation monitors are also an option for low count rate situations. Fission chambers have an advantage of a very high signal and therefore a very strong $\gamma$-ray rejection.

It is crucial that the selected neutron beam monitor fulfills the instrument requirement ~\cite{bateman2002high,pietropaolo2009single, de2003neutron, Belushkin2008}. 
For instance, a beam monitor tested at NOP beamline in J-PARC showed a variation in the neutron detection efficiency which exceeds the tolerable level  for high precision measurement at the neutron lifetime measurement~\cite{Ino2014}. 
Several beam monitors will be used at ESS, some will be realized for ESS~\cite{Croci2014,albani2016evolution,croci2013gem} others will be manufactured by different suppliers.
The aim of this work is to characterize seven neutron beam monitors, manufactured by several companies  and  to compare their properties  to be categorized  for various applications.  

\section{Detection techniques}
Gas proportional counters have dominated the world of thermal neutron detection because of their robust design and high detection efficiency and low cost compared to solid-state detectors. Their operation is based on the absorption of thermal neutrons in the detector-converting medium. Lacking charge, thermal neutrons cannot interact with matter through Coulomb interaction, which is the dominant mechanism through which charged particles  lose their energy. Therefore mechanisms for detecting thermal neutrons are based on indirect methods where neutrons interact with specific atomic nuclei to produce  energetic ions. The most commonly used neutron interactions for thermal neutron detection are
summarized in  Table~\ref{tab:rx}. 

\begin{table}[H]
  \centering
  \caption{Nuclear reactions for thermal neutron detection}
  \label{tab:rx}
  \begin{ruledtabular}
  \begin{tabular}{cccc}
    Reaction &Isotope& Cross  & Q-value \\
     &abundance (\%)&  section (barn) &  (MeV)\\
     &                         &  at 1.8 ~ \AA     &   \\
    \hline
       $^3$He(n,p)$^3$H        & 0.014  &5330   & 0.76 \\ 
       $^{10}B(n,\alpha)^7$Li  & 19.9   &3838   & 2.31(94\%) \\
         &  && 2.79(6\%) \\
       $^6 Li(n,\alpha)^3$H        &   7.6   & 947   & 4.78 \\
       $^{14}$N(n,p)$^{14}$C & 99.63  &1.91  &0.62\\
       $^{235}$U(n,3n)ff         &0.72    &680.9  & $\approx$200\\

  \end{tabular}
   \end{ruledtabular}
 \end{table}

\begin{figure*}
\includegraphics[width=170mm]{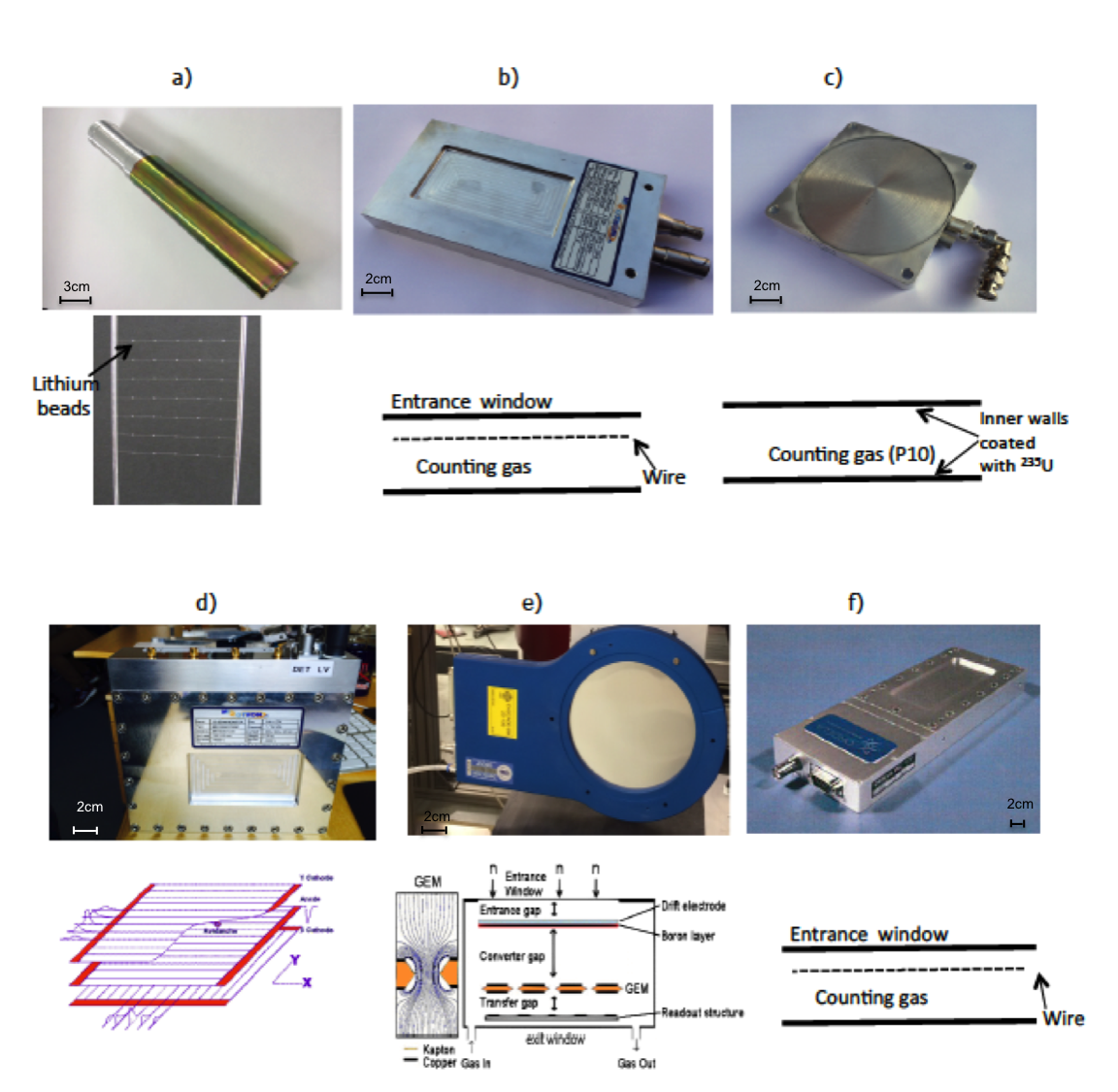}
\caption{Photographs and diagrams of the neutron beam monitors tested in this work: a) scintillator from Quantum Detectors UK, b) MWPC from Mirrotron-0D, c) fission chamber form LND, d) MWPC from Mirrotron-2D, e) GEM from CDT f) MWPC from ORDELA }
\label{BM_pic}
\end{figure*}

The  beam monitors under study are shown in figure~\ref{BM_pic}.  Their specifications are summarized in Table~\ref{specification}. These monitors can be categorized depending on their design and their equipped electronics into two main categories: position sensitive and non-position sensitive. The multi-wire proportional chambers (2D-MWPC) from Mirrotron  ~\cite{MIR} and the gas electron multiplier (GEM) ~\cite{Sauli, sauli1997gem} from CDT  ~\cite{CDT} are  position sensitive beam monitors, while other beam monitors are area-integrating, i.e. flux monitors.  

All monitors studied here detect discrete neutron events, rather than an integral charge proportional to flux and all are sealed except the GEM monitor from CDT. This monitor should be filled with a gas mixture of Ar/CO2 with a flow rate of 5-10 sccm. To determine the exact position of the neutron incidence, the detector includes a position sensitive  readout electrodes. 

\begin{table*}
  \caption{ Specifications of the  tested neutron beam monitors }
 \label{specification}
 \begin{ruledtabular}
 \begin{tabular}{cccccccc}

     Monitor         &  MWPC    &      MWPC  &   MWPC   &      2D-MWPC   &      GEM              &    Scintillator        & fission chamber   \\
    Manufacturer & ORDELA  & ORDELA  &  Mirrotron  &   Mirrotron      &       CDT                 & QD                      & LND\\
    
     \hline
     \hline
     \\
    
 active element   & $^{3}$He & $^{14}$N  & $^{3}$He & $^{3}$He       & $^{10}$B    & $^{6}$Li    &$^{235}$U \\ 

   partial pressure &  6.07 &  81.06  & 6.5   &  0.4    &  -   & - &  \\ 
   total  pressure &     -       &     -        &  -      &     -      & 1000    & .. & 1013.2 \\  
         (mbar)       &         &               &         &             &         &      & \\ 
filled gas & $^{3}$He+$^4$He  & N+CF$_4$ & $^{3}$He+CF$_4$  & $^{3}$He+CF$_4$& Ar+CO$_2$& ..&P10\\
                &  +CF$_4$               &                    &                                &                               &                 &   & \\

 bias voltage(V)& 850 & 850& 1300 &anode:-3500  &-1000 & 650&300 \\    
        &  & &  & drift:1500& & & \\ 
   \\     
active area (mm)  &  114x51          &114x51 &100x50  & 100x50 &   100x100       &28x42 &     100x100 \\    
 window thickness (mm)&	2&	2&	1& 1&	0.1& 0.1&1
       
  \end{tabular}
  \end{ruledtabular}
\end{table*}

\subsection{Multi-wire proportional chambers }
The monitors from Mirrotron and ORDELA are multi-wire proportional chambers (MWPC) with low detection efficiency. Their entrance window is a thin aluminum membrane (around 1 mm thick) to minimize the neutron attenuation coefficient. The two ORDELA models ~\cite{ORDELA}  are filled with $^{3}$He+$^{4}$He+CF$_4$ and  with $^{14}$N+CF$_4$  respectively, while the Mirrotron models are filled with $^{3}$He+CF$_4$. 
The CF$_4$ gas is added as a stopping and quenching gas to limit the proton and triton ranges \cite{Christophorou1982} and to control the amplification. 

The  2D-MWPC from Mirrotron is a position sensitive monitor with a delay line readout. It is a chamber filled with a converter gas ( $^{3}$He+CF$_4$) for thermal neutron detection and consists of one anode frame and two cathode frames wired with gold-plated tungsten wire  (1000 pixels). When a neutron is captured in the converter gas, electron-ions pairs are created. Then the electrons are accelerated toward the wire electrodes by the applied electric field. It is possible to determine the position of interaction by the integrated electronics. For this monitor the position encoding is based on delay lines ~\cite{orban2015serial}. The set of cathode wires is connected to a delay line read-out system. The delay time differences are measured for each set of cathodes with time-to-digital converter and these digital signals give an XY position encoded output. Beside the provided XY signals there is the anode signal which is the summed signal, this is used to obtain the pulse height spectra (PHS).   

\subsection{Scintillator}
Scintillator thermal neutron detectors are common as a Li-based thermal neutron detector this is because a Li-containing gas is not available. The principle of detection relies on the emission of light photons due to the interaction of radiation with the scintillation material \cite{Rhodes2004243}. Usually a photomultiplier is used to convert the light output of a scintillation pulse into a corresponding electrical signal. A scintillator detector from Quantum Detectors UK ~\cite{ISIS}, consisting of a 2-dimensional array of neutron active scintillator beads (35 beads each is 0.25 mm$^3$) made of lithium silicate glass has been used in this work. The active area is 28x42 mm$^2$.

\subsection{Fission chamber}	
 Neutron-induced fission reactions can also be used for thermal neutron detection.  In this case, extremely low background rates can be achieved thanks to the large Q value ($\approx$200 MeV) of the reaction. The most popular form of a fission detector is an ionization chamber coated with a layer of a fissile deposit such as $^{235}$U. Generally the fissionable materials used  in a fission chamber are alpha emitters. Consequently, fission chambers have undesired pulse rates from the alpha background. This can be easily discriminated based on the pulse amplitude since the energy of an emitted alpha  is 10 times smaller than the average energy of fission fragments.  Typically the operating voltage of a fission chamber is between 50 and 300 V. The principle of neutron detection in a fission chamber relays on the produced fission fragments that deposit their energy in the active volume, which is normally filled with argon. The  pressure of the gas inside the chamber ensures that the range of fission fragments within the gas does not exceed the small dimensions of the detector.
 The fission chamber used in this work was purchased from LND ~\cite{LND}. It contains 13 mg of $^{235}$U and  is filled with P10  gas (90\% argon and 10\% methane). The detection depth is 19.1 mm.

\subsection{Gas Electron Multiplier} 
It is also possible to use a solid neutron converter with a gas readout. For the GEM beam monitor from CASCADE a 0.8 $\rm\mu$m  thick  boron layer is used as a neutron converter layer. The thickness of the boron layer is chosen here  for measurement purpose. However, a thinner layer will be selected for a full implementation i.e. around  100 nm. 
This layer is coated on the drift electrode. The created charged products (alpha  and triton) deposit their energy in the counting gas creating e-ion pairs along their track. The electrons then drift toward the GEM multiplier foil that is made of 50 $\rm\mu$m  of Kapton substrate sandwiched between 5 $\rm\mu$m  thick copper claddings on either side. The detector is supplied with its readout system. Each electrode of the readout structure is connected to one channel of a highly integrated ASIC, which contains for each channel a charge sensitive pre-amplifier, shaper and discriminator.

\section{Measurements Methods} 

The gamma-ray sensitivity of a neutron beam monitor is an important criterion in its selection. This is because it is important to discriminate between the  background radiation which is mainly gamma and the  neutrons from the beam. To understand the degree of neutron to gamma discrimination, the beam monitors were tested using neutron and gamma sources at the Source Testing Facility, Lund University. The used neutron source is Be-based which emits neutrons up to about 11MeV with a maximum intensity in the 3-5 MeV range \cite{scherzinger2016comparison}. In order to thermalize the emitted neutrons, the source is placed in a moderator made of polyethylene. For gamma sensitivity measurements different gamma sources ($^{60}$Co, $^{57}$Co, $^{133}$Ba) at 2 cm from the monitor were used   in order to evaluate the sensitivity of the studied monitors to low, medium and high-energy gamma. The corresponding energies and intensities of these sources are summarized in Table~\ref{gammasource}. 

The gamma spectra was compared with the background measurement which means measurement  in the absence of gamma and neutron sources. PHS are obtained as counts versus channel number. The channel numbers correspond to the energy deposited by the detected radiation in the active depth of each monitor.

\begin{table}[htbp]
  \centering
  \caption{The gamma sources used with their corresponding energy and intensities}
\label{gammasource}
\begin{tabular}{cccc}
Isotope & Current activity   &   Energy    &   Intensity      \\
 &  (MBq)  &    (keV)   &    (\%)       \\
\hline
 & & & \\
     
 $^{60}$Co& 15.58 & 1173.23 & 99.85 \\ 
  & &1332.48& 99.98\\
  $^{57}$Co & 21.34 & 122& 85.6\\  
   & & 136&10.7 \\ 
    $^{133}$Ba & 3.12 & 384& 9\\
     & & 356& 62\\
     & & 303& 18\\  
     & & 276& 7\\ 
    & & 80& 36\\
    && 30-35& 117
   
\end{tabular}
\end{table}
For measuring the ORDELA and Mirrotron monitors  and the fission chamber we used the ORTEC 142-PC preamplifier. The preamplifier output was connected to a                                                                                                                              shaping amplifier. The output signal is sent to a multichannel analyzer to obtain the pulse height spectra (PHS). Even though this type of the ORDELA monitors is supplied with preamplifier and discriminator, the anode signal was directly connected to the ORTEC 142-PC preamplifier for a more detailed analysis.  
The nominal high voltage (HV) up to 3500V  is provided by a HV power supply CAEN 1470.

The scintillator beam monitor was supplied with its discriminator  electronics. The discriminator  processes the electrical signal producing two outputs: analogue and digital signal. The digital output appears whenever the signal amplitude is greater than the threshold. This is set so that the gamma part of the spectrum is rejected. However, for a more detailed analysis the analogue signal was used. Then this signal is sent to the Multi-channel analyzer (MCA) to obtain the pulse height spectra. The measuring time is 10 minutes unless  stated otherwise.

\begin{figure}[H]
\centering
\includegraphics[width=70mm]{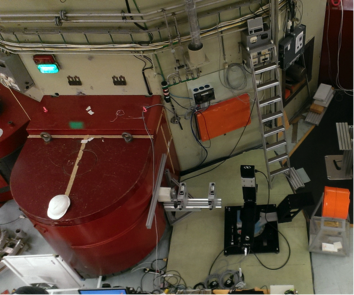}
\includegraphics[width=70mm]{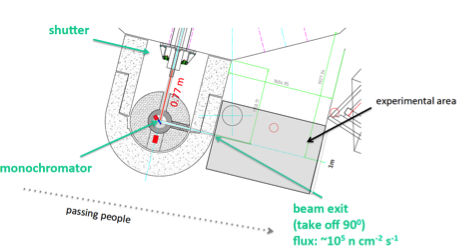}
\caption{The setup used for thermal neutron detection at R2D2 beam line at IFE-Norway }
\label{R2D2setup_pic}
\end{figure}

For efficiency  measurements we used the R2D2 beam-line at the JEEPII reactor at IFE, Norway.  This beam comes from the reactor through a shutter as shown in Figure~\ref{R2D2setup_pic}. The collimator in the shutter is made of Cd layers and has a 36 $^{\circ}$  divergence. The beam is then monochromated by a composite Ge wafer monochromator. The beam size at the tested monitors is controlled by two JJ X-Ray IB-C80-AIR slits with borated aluminum blades which are placed before the monitors. The expected neutron flux at the center of the beam is of the order of 10$^5$ n/ s.cm$^2$ thermal neutrons. The wavelength of the neutron beam can be adjusted from 0.88~\AA \ to 2.4~\AA \ by the rotation of the monochromator. For high statistics, all the measurements were performed at 2.4~ \AA  \ neutron beam. 

\begin{figure}[H]
\centering
\includegraphics[width=80mm]{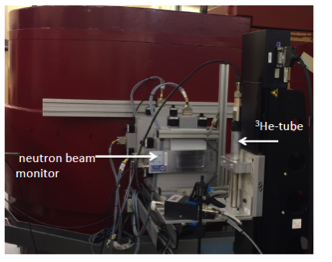}
\caption{Setup used at  R2D2 beam for efficiency and attenuation measurements}
\label{eff-setup}
\end{figure}

Efficiency was measured using a 1-inch stainless steel $^3$He-tube (Tube RSP40810227 manufactured by GE-Reuter Stokes). The tube was operated at a $^3$He pressure of 10 atm, and a voltage of 1400 V. The walls of
the tube are made of stainless steel and have a thickness of 0.508 mm.  
The efficiency setup used is shown in Figure~\ref{eff-setup}, where both the He-tube and the monitor are irradiated with the same beam successively.  It is important to make sure that the beam is narrow enough so that it is fully measured by the He-tube and its intensity is adjusted so no saturation occurs in the tube. 
For the beam monitors with very low detection efficiency one should measure for long time to accumulate enough statistics.

\section{Gamma Sensitivity}

\begin{figure}
\includegraphics[width=70mm]{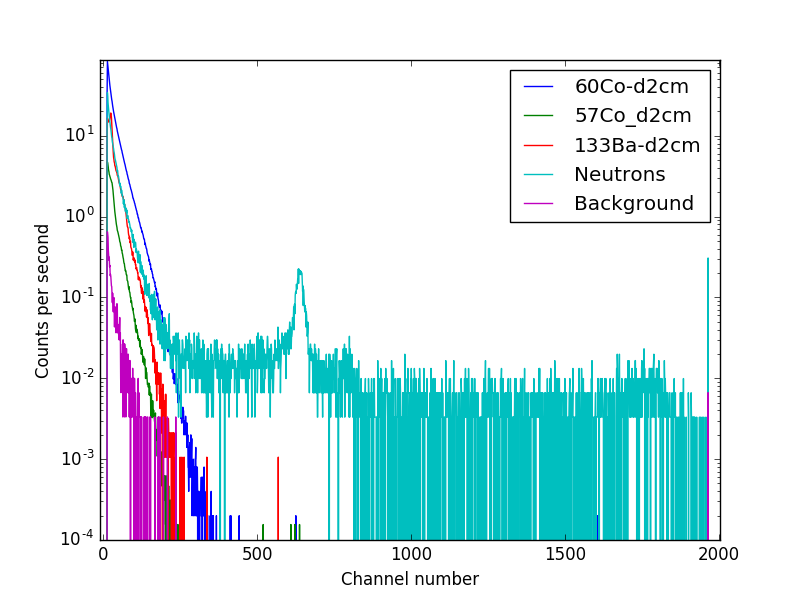}
\caption{Pulse height spectra measured by the Mirrotron monitor using a neutron source and using three gamma sources separately ($^{60}$Co, $^{57}$Co, $^{133}$Ba) (1 channel number corresponds to 1.2 keV  ). 
\label{MIR_1d_gamma}}
\end{figure}

\begin{figure} [H]
\includegraphics[width=70mm]{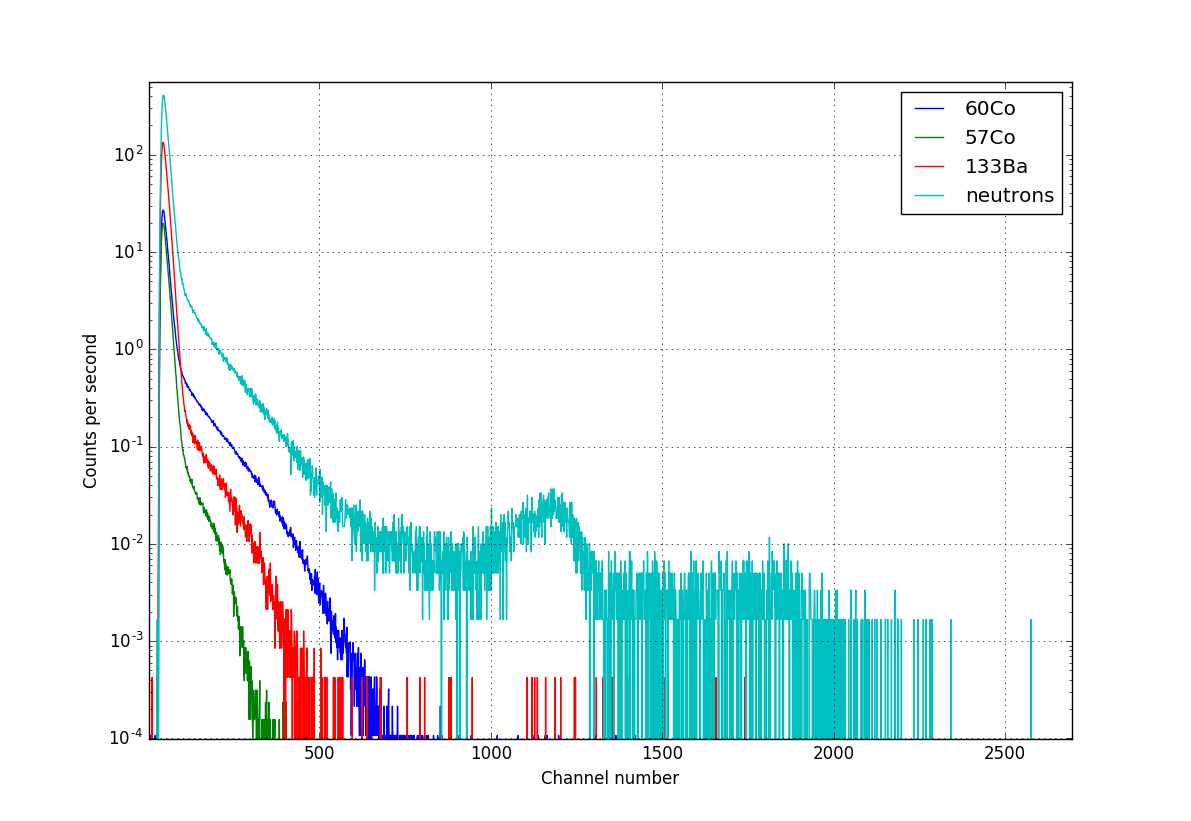}
\caption{Pulse height spectra measured by the Mirrotron-2D monitor using a neutron source and using three gamma sources separately ($^{60}$Co, $^{57}$Co, $^{133}$Ba) (1 channel number corresponds to 0.6 keV  ). 
\label{MIR_2d_gamma}}
\end{figure}

\begin{figure}[H]
\includegraphics[width=70mm]{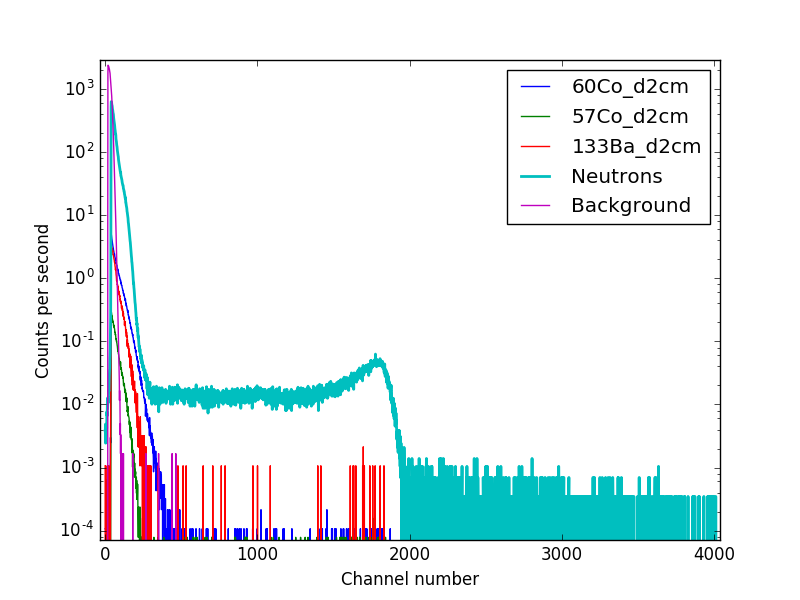}
\caption{Pulse height spectra measured by the ORDELA-He monitor using a neutron source and using three gamma sources separately ($^{60}$Co, $^{57}$Co, $^{133}$Ba) (1 channel number corresponds to 0.4 keV  ). }
\label{ORDELA_He_gamma}
\end{figure}

\begin{figure}[H]
\includegraphics[width=70mm]{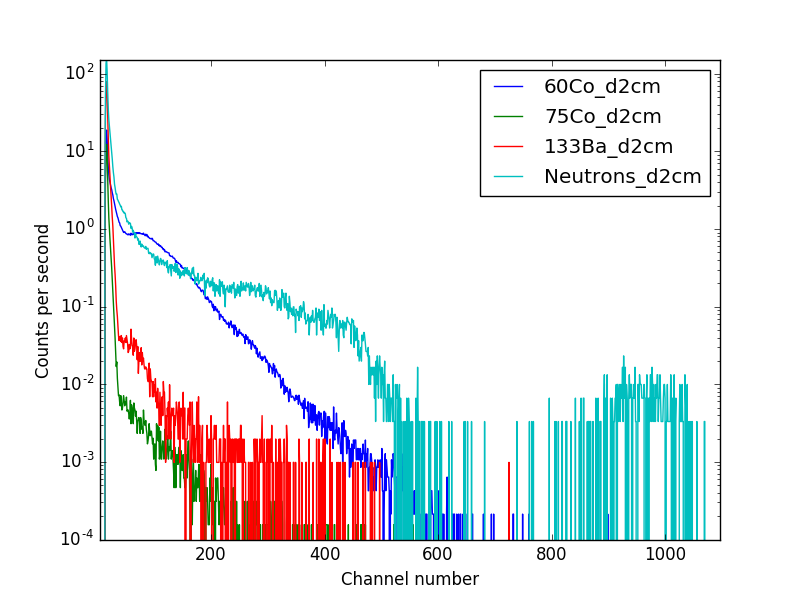}
\caption{Pulse height spectra measured by the scintillator monitor using a neutron source and using three gamma sources separately ($^{60}$Co, $^{57}$Co, $^{133}$Ba) (1 channel number corresponds to 5 keV  ).}
\label{ISIS_gamma}
\end{figure}

\begin{figure}[H]
\includegraphics[width=70mm]{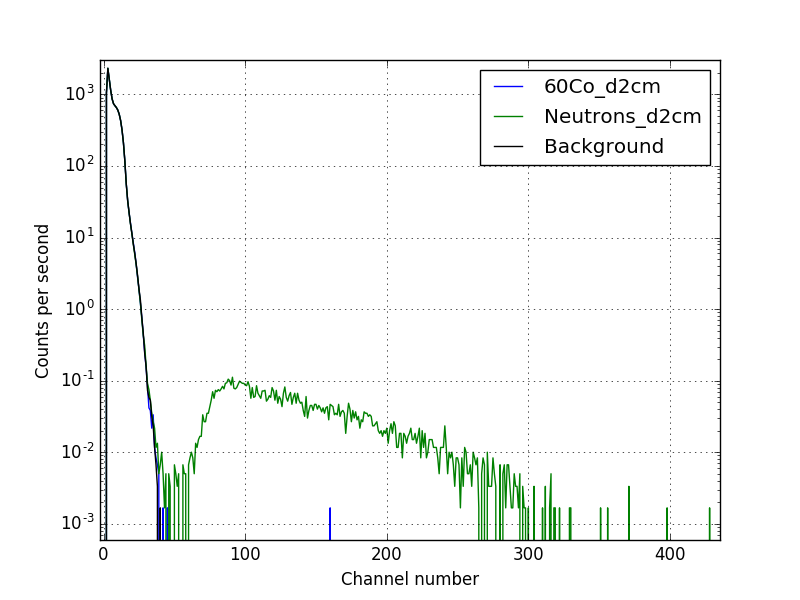}
\caption{Pulse height spectra measured by the LND fission chamber using a neutron source and using  $^{60}$Co (1 channel number corresponds to 2.1 keV  ).}
\label{LND_gamma}
\end{figure}

\begin{figure}[H]
\centering
\includegraphics[width=78mm]{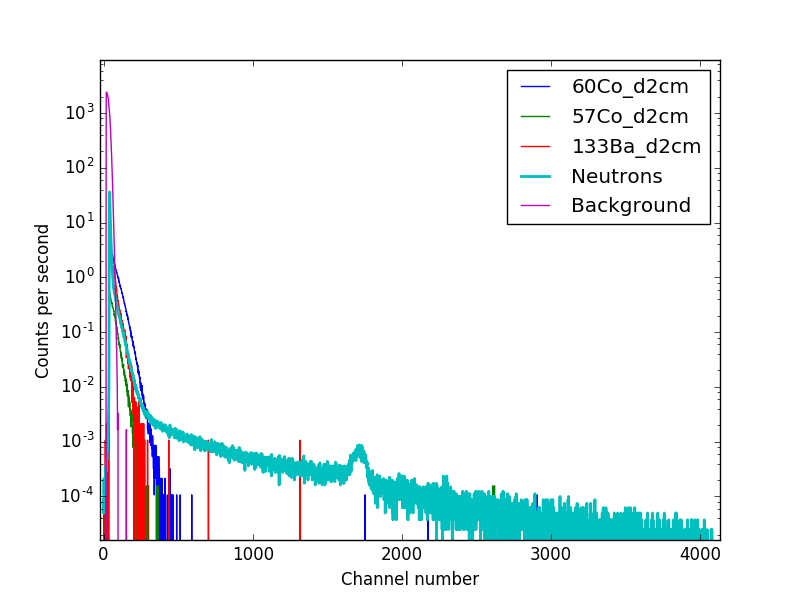}
\caption{Pulse height spectra measured by the ORDELA-N monitor using a neutron source and using three gamma sources separately ($^{60}$Co, $^{57}$Co, $^{133}$Ba) (1 channel number corresponds to 0.4 keV  ).
\label{ORDELA_N_gamma}}
\end{figure}

The PHS obtained by the Mirrotron monitor (BM-100X50) filled with $^3$He+CF${_4}$ are shown figure~\ref{MIR_1d_gamma}. Neutron detection results in a peak in the higher channel numbers (channel number 650), while gamma contributes to the lower channel numbers (lower than 360 channel number). There is a valley between the gamma and neutron detection which allows  gamma discrimination. As the energy of gamma increases, a higher energy deposition is obtained in all the monitors. This is seen by a greater energy deposited (higher channel number) by the $^{60}$Co compared to $^{133}$Ba and $^{57}$Co as shown in figure~\ref{MIR_1d_gamma} to  figure~\ref{ORDELA_N_gamma}. An approximate normalization of the channel number is stated in the caption of figure ~\ref{MIR_1d_gamma} to~\ref{ORDELA_N_gamma}. This normalization is based on the position (channel number) of each peak in the PHS. 

Having the same converting gas ($^3$He), the PHS features of the Mirrotron, Mirrotron-2D and the ORDELA-He look similar. However, the ORDELA produces a much clear and sharp edge, while the Mirrotron has a clearer narrow peak.

The PHS of the scintillator  monitor is shown in figure~\ref{ISIS_gamma}.
It is clear that the gamma region extends from low channel number up to channel number 600. The neutron peak is well resolved at higher channel number (between 800 to 1000) which allows a good discrimination between gamma and neutrons . The neutron detection is revealed by the presence of a neutron peak around channel number 970.

On the other hand, due to the high Q value ($\approx$200 MeV) , the fission chamber with $^{235}$U as a detection medium shows good gamma discrimination as shown in figure~\ref{LND_gamma}. No gamma signals from any source could be detected in the fission chamber with the experiment used. All the  spectra are identical in the low-energy region. This part of the spectrum comes from alpha decays of $^{235}$U which are few up to 10 MeV. Thus a very clean discrimination can be accomplished using fission chambers.

Finally, nitrogen can also be used as a neutron converter gas when very low neutron detection efficiency is required. The neutron detection is based on charged particles created from   $^{14}$N(n,p)$^{14}$C reaction. Due to the low cross section of N with neutrons and the low efficiency of this monitor longer time measurements (22 hours) was required to obtain statistics. As shown in figure~\ref{ORDELA_N_gamma}, the neutron detection is revealed by the presence of the peak at high channel number. Gamma detection is revealed by the counts  from all the sources in to the low-energy region.   


The PHS  was not possible to obtain for the $^{10}$B-based GEM monitor from CDT, based on the electronics provided. Thus the 2D-image was used for the  gamma sensitivity  measurements showing almost no gamma detection even when the monitor was tested at the lowest threshold level. This could be attributed to the threshold level  which is already set in a way to reject the gamma. This gamma insensitivity has been also observed in other GEM-based neutron detectors~\cite{croci2013gem}.

\begin{figure}[H]
\centering
\includegraphics[width=90mm]{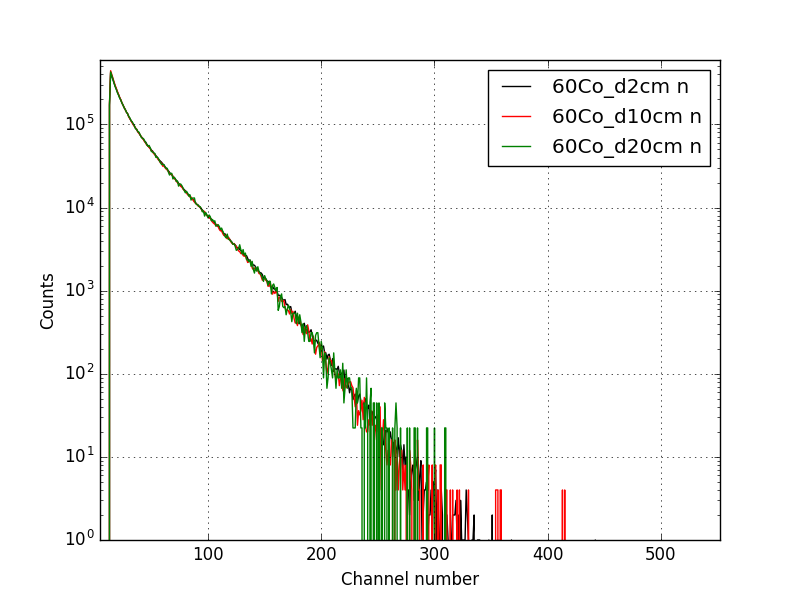}
\caption{Pulse height spectra measured by the Mirrotron monitor  using  $^{60}$Co at three distances.}
\label{MIR-1d_gamma_pileup}
\end{figure}

To better understand if there is a pile-up effect in these monitors, they have been measured  at several distances with respect to the used  gamma source. Figure~\ref{MIR-1d_gamma_pileup} is a an example of this study. The PHS are normalized with respect to their corresponding solid angle. It is shown that with the used gamma sources there is no significant effect.

\begin{figure}[H]
\centering
\includegraphics[width=100mm]{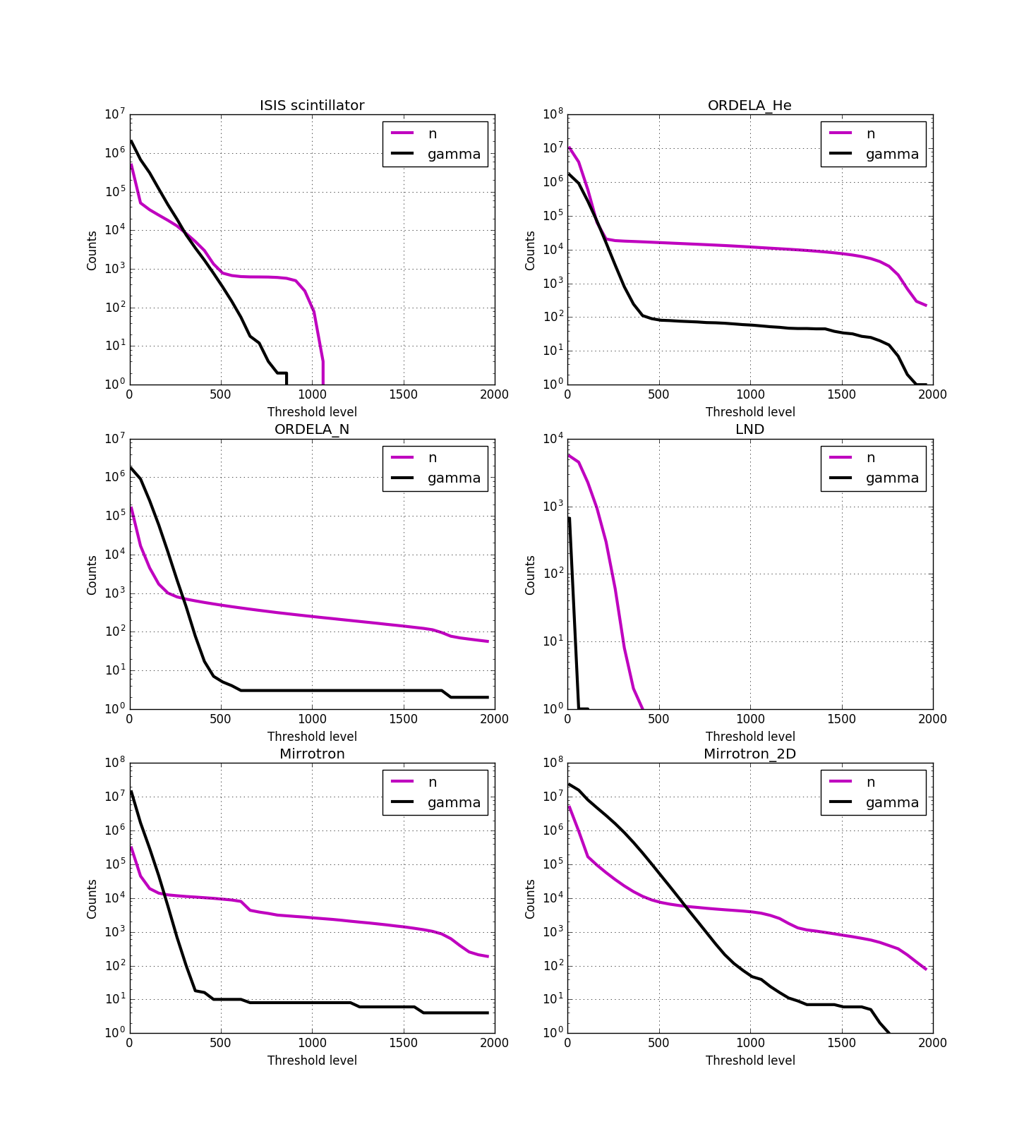}
\caption{Gamma and neutron counts using $^{60}Co$ and Be-based neutron sources respectively. These counts are collected in 10 min as function of the set threshold.}
\label{gamma_thr}
\end{figure}

Figure~\ref{gamma_thr} shows  the  gamma and neutron counts from the measurements where $^{60}$Co or Be-based neutron source was used respectively. These counts are collected in 10 minutes as function of the threshold level. Gamma and neutron counts decrease with the increase in the threshold level. For lower gamma sensitivity the threshold level should be set in a region which maximizes neutron counts and minimizes gamma counts. This differs from one monitor to another based on the monitor gamma sensitivity. In addition, a correctly chosen threshold level allows a high gamma rejection ~\cite{khaplanov2013investigation}. For instance the fission chamber shows the lowest gamma sensitivity where the region of high gamma counts is very  narrow allowing high neutron counts with negligible gamma counts.

\section{Neutron Detection}

\begin{figure}[H]
\centering
\includegraphics[width=70mm]{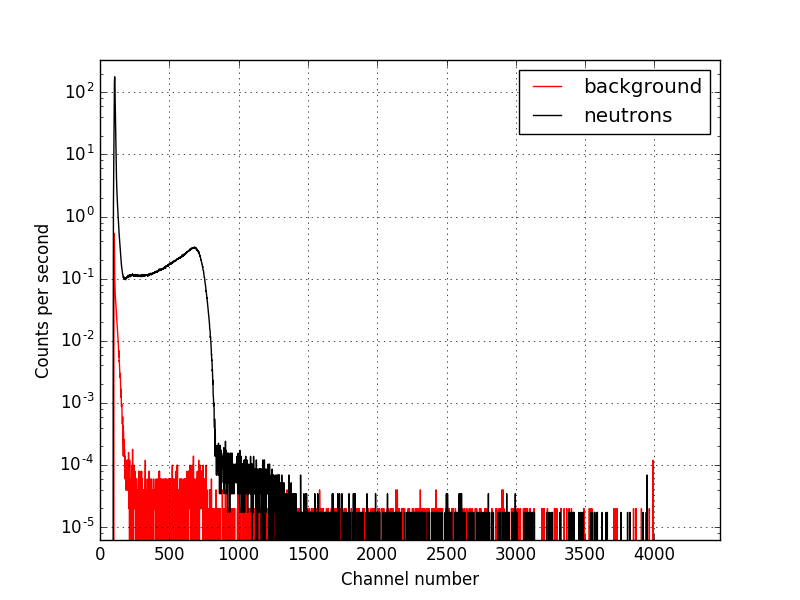}
\caption{Pulse height spectra measured by the ORDELA-He monitor  using neutron beam line at 2.4 ~\AA  (1 channel number corresponds to 1.1 keV  ).
\label{ORDELA_He_n}}
\end{figure}

\begin{figure}[H]
\centering
\includegraphics[width=70mm]{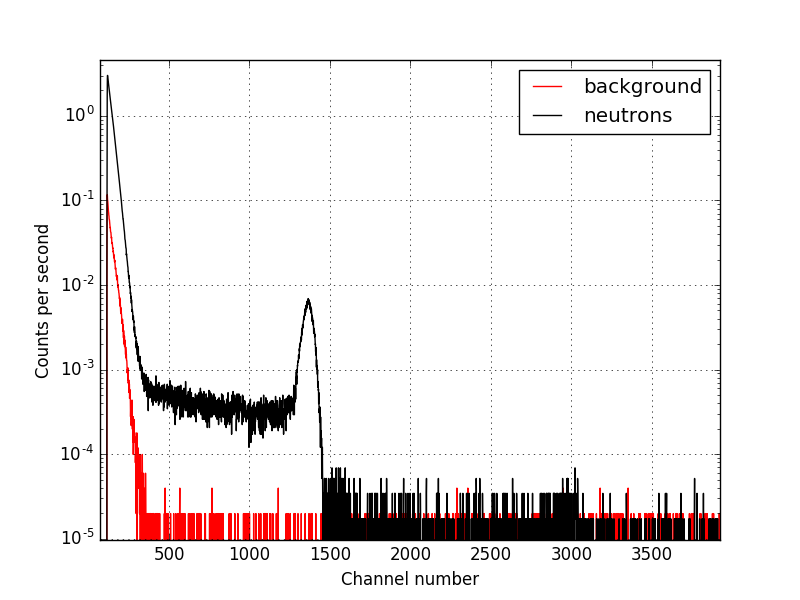}
\caption{Pulse height spectra measured by the ORDELA-N monitor using neutron beam line at 2.4 ~\AA (1 channel number corresponds to 0.45 keV  ).
\label{ORDELA_N_n}}
\end{figure}

\begin{figure}[H]
\centering
\includegraphics[width=70mm]{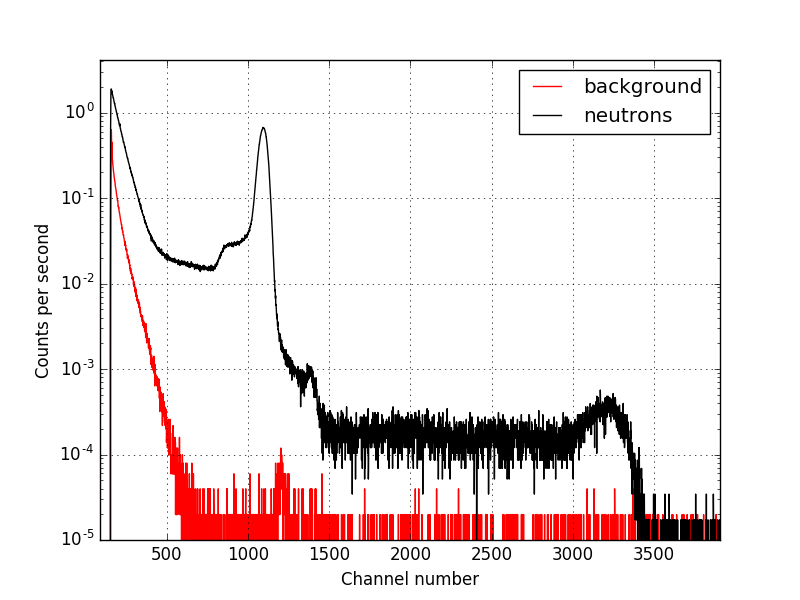}
\caption{Pulse height spectra measured by the Mirrotron monitor using neutron beam line at 2.4 ~\AA (1 channel number corresponds to 0.7 keV  ).
\label{MIR1d_n}}
\end{figure}

\begin{figure}[H]
\centering
\includegraphics[width=75mm]{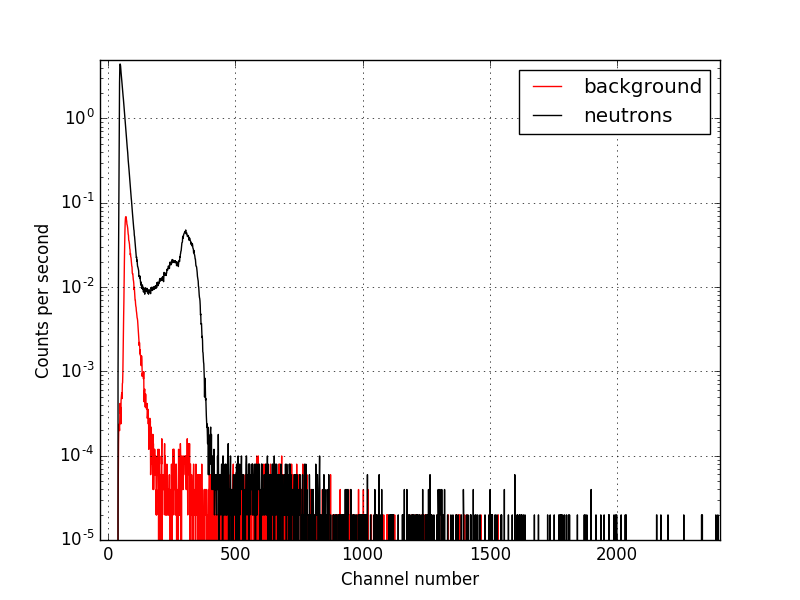}
\caption{Pulse height spectra measured by the Mirrotron-2D monitor using neutron beam line at 2.4 ~\AA (1 channel number corresponds to 2.5 keV  ).
\label{MIR2d_n}}
\end{figure}

Figure~\ref{ORDELA_He_n} to ~\ref{MIR2d_n} show the PHS of different beam monitors using the neutron beamline at IFE and background when the beam is off. The neutron detection is revealed by the neutron peak in all the spectra. From such spectra  information  regarding efficiency were extracted. The neutron detection is revealed in all the spectra by the presence of a peak while gamma counts contribute to the lower channel number.
In the Mirrotron and the ORDELA-He monitors the peak is around the 760 keV while for the ORDELA-N the peak is at 620 keV.

\subsection{Efficiency, attenuation and scattering measurements}

\begin{figure}[H]
\includegraphics[width=60mm]{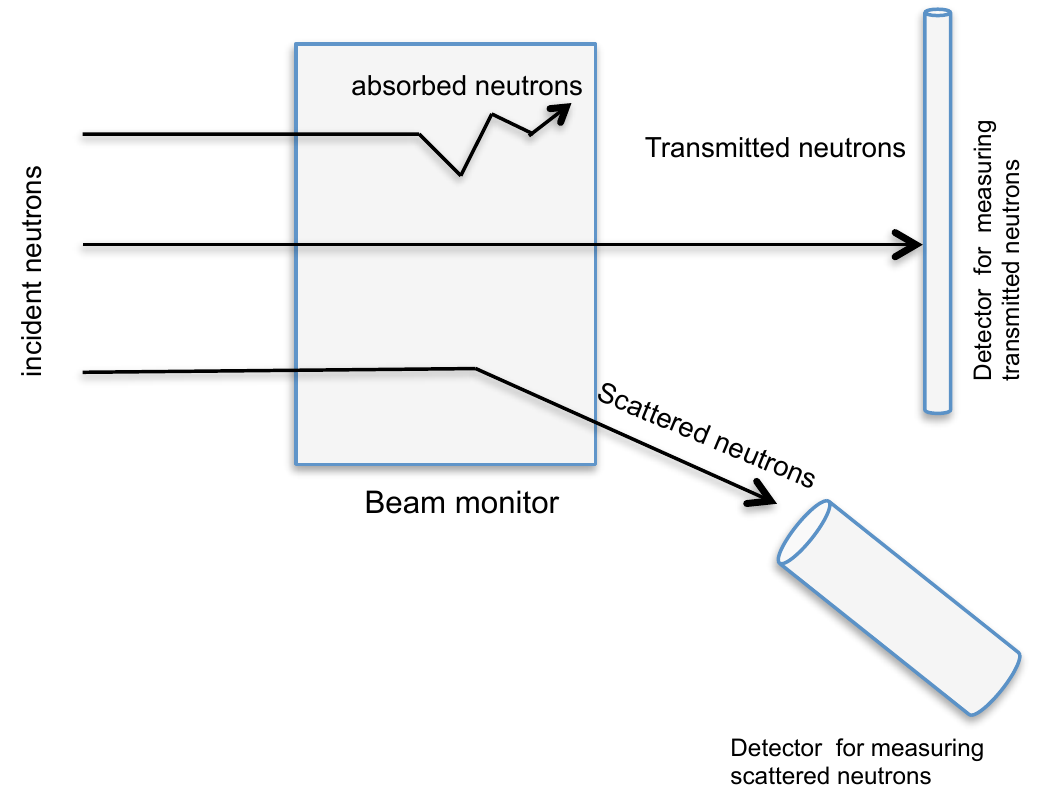}
\caption{Schematic presentation showing the setup used for measuring transmission and scattering of neutrons, the detector used for transmission is a $^3$He-tube and a $^6$Li-glass scintillator from SCIONIX used for measuring scattering }
\label{att-tran-sact}
\end{figure}

 The efficiency of each monitor is measured by dividing the total number of counts above the noise level and the total number of neutrons in the beam which is estimated from the total counts in the  $^3$He-tube counts ( 96 \% efficiency). The efficiency has been evaluated at a $\lambda$~=~2.4~\AA. 
These results are  summarized in Table~\ref{eff-att}. 
As it is important to monitor the neutron beam at each instrument then it is critical that the used beam monitor allows the passage of a neutron beam without affecting it. Since there is always a probability of absorbing or scattering a neutron by the monitor and its housing material it is important to study their corresponding attenuation coefficients. 
This is illustrated in figure~\ref{att-tran-sact} where the transmitted neutron flux  (T) is equal to the incident flux (F) minus the attenuated neutrons which is  due to  scattering (S) and absorption (A).
\begin{equation}\label{eq:att}
 T=F-A-S
\end{equation}

\begin{equation}\label{eq:att}
 T/F= transmission factor
\end{equation}

For these  measurements the same setup as for the efficiency was used except that the monitor was placed and removed in front of the $^3$He tube in order to measure loss of counts due to the monitor. This allows measuring the fraction of neutrons attenuated by the measured monitor. The scattered neutrons were measured by a $^6$Li-glass scintillator detector from SCIONIX with 99.98 \% efficiency. The beam stability was monitored by a monitor from Canberra while performing these measurements to ensure the stability of the neutron beam.
 The total attenuation, efficiency and scattering percentages are summarized in table~\ref{eff-att}.  The statistical error is relatively low compared to the measured attenuation and efficiency percentages. For scattering measurement the main source of standard deviation is due to the solid angle calculation these values are summarized in table~\ref{eff-att}.

\begin{table*}[htp!]
  \centering
 \caption{ The measured efficiency and attenuation by all the monitors using the R2D2 beam-line at  2.4 \AA }
 \label{eff-att}
 \begin{ruledtabular}
 \begin{tabular}{ccccccc}

monitor&  measured & calculated  & supplier  & measured &  calculated  & scattering \\
& efficiency at 2.4 ~\AA (\%)& efficiency (\%) &  efficiency (\%)  at 1.8 ~\AA & attenuation (\%)&   attenuation (\%)&  (\%)\\
      \hline
   
ORDELA-He & 0.12 & 0.13  & 0.1  & 4.5 & 4.05 &3.9 $\pm$ 0.4  \\
   
ORDELA-N     & 2.1 x 10$^{-3}$ & 1.3 x 10$^{-3}$& 1 x 10$^{-3}$&4.4& 4.05 & 3.8 $\pm$ 0.4\\
Mirrotron         &  0.11               & 0.13 & 0.1 & 2.5&2.05 &4 $\pm$ 0.9\\
Mirrotron- 2D  & 1.5 x 10$^{-2}$         &   1.7 x 10$^{-2}$     &1.5 x 10$^{-2}$&7.3& 4.5 & 9 $\pm$ 1 \\
CDT                & 3              &   4                           & 3                     & 10.6&     6.9 & 10.3 $\pm$ 0.7 \\
Scintillator      & 0.052           &       0.012    & 0.01   & 0.49& 0.1&  0.74 $\pm$0.2 \\
Fission chamber & 0.01      & 0.021& 0.01 & 3.87& 2.4 & 3.8 $\pm$ 0.3\\

 \end{tabular}
\end{ruledtabular}
\end{table*}

As the GEM monitor from CASCADE has a relatively thick layer (0.8 $\rm\mu$m) of boron, it shows the highest efficiency. The relatively high efficiency of this monitor is chosen to reduce the measuring time.
The ORDELA monitors show different efficiency depending on the used gas cross-section. For instance the $^{14}$N cross section is several orders of magnitude lower than $^{3}$He making the latter more sensitive to neutron detection.  The ORDELA and Mirrotron monitors filled with $^{3}$He have similar parameters and comparable efficiencies.The relatively high efficiencies for the monitors of each type have been selected to expedite the measurement times.
The measured efficiency has been compared to the calculated efficiency based on the supplier information regarding the gas pressure and the depth of the counting volume. This calculation is based on a simple consideration of cross sections. This reveals comparable results with the measured ones. The equation used for  calculating the efficiency is given below:

\begin{equation}\label{eq:eff}
 \varepsilon  =1-exp(-\mu\lambdaρχ. \rho x)
\end{equation}

where
$\varepsilon$  is the efficiency, $\mu$ is the gas absorption factor for the detector gas  {(atm.cm .$\lambda$)}$^{-1}$,  $\lambda$ is the wavelength in $\AA$, $\rho$ is the absolute counting gas pressure and  
x is the depth of the counting volume. 

The high level of attenuation shown by all the monitors except the scintillator monitor is striking. For instance, for the ORDELA monitors the neutron beam has to cross the 0.2 cm Al window on both sides and the 1.3 cm sensitive depth made of either He or N. The Mirrotron-2D showed three times higher attenuation compared to the Mirrotron monitor. This could be attributed to the  Kapton  foil  and the thicker window used inside the detector. This 2D beam monitor type could be used  for beam diagnostic purposes removing it from the neutron beam when it is not in use.
The scintillator  monitor shows the lowest attenuation percentage. This could be explained by a thinner aluminum window (100 $\rm\mu$m ) window. The LND depth is 19.1 mm and is coated with 13 mg of $^{235}$U it attenuates the beam by 3.8\%.  The difference between the measured and the calculated attenuation in the LND fission chamber may be attributed to the way the uranium is coated on the Al plate where additional materials might be used to make the adhesion possible. 
The ORDELA monitors attenuates the beam by about 4 \% which is due to the 4 mm Al window thickness on both sides. The high attenuation factor in most of the monitors is mainly due to the thickness of the window. Thinner Al windows  (less than 1 mm) should be used for lower attenuation effects.

\subsection{Linearity}

It is important that a used beam monitor measures  the rate in the neutron beam with a high precision. Normally the response of a beam monitor should be linearly correlated to a change in the neutron beam. For studying the linearity response of these monitors as function of the change in the  beam flux. Several plastic plates (one plate thickness is around 3 mm) were placed close to the beam-line exit  and 1 m away from the beam monitor used as neutron attenuators.  The attenuation in the neutron intensity by using several plates were measured by the $^3$He-tube.  The beam is attenuated by about 50\% by using one plate. 
 
\begin{figure}[H]
\centering
\includegraphics[width=80mm]{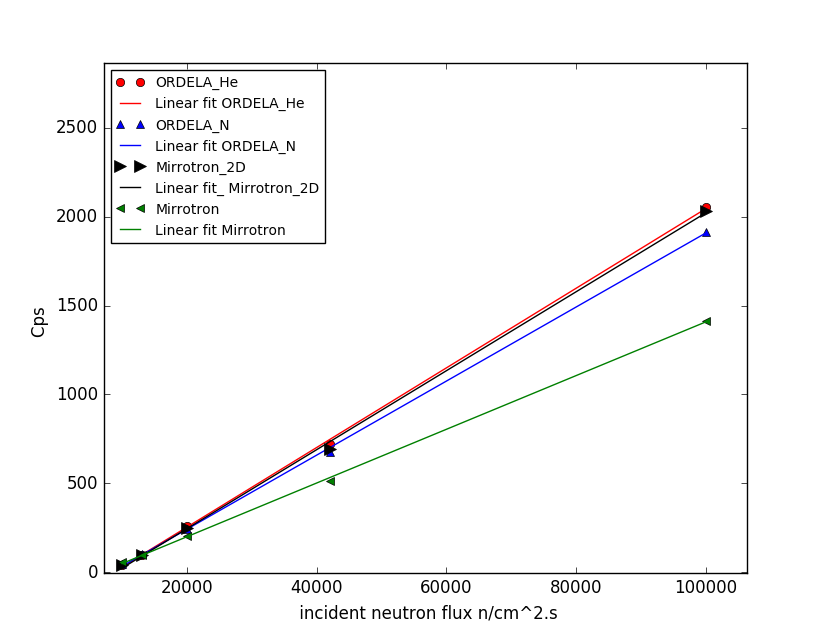}
\caption{Neutron counts measured by different monitors versus the variation in the incident neutron flux.  Counts from ORDELA-N and Mirrotron-2D monitors were multiplied by a factor of 100 and 10 respectively.   }
\label{linearity-2}
\end{figure}

The neutron counts from different beam monitors as function of the change in the incoming neutron flux are shown in  figure ~\ref{linearity-2}.  The results show that there is no measurable deviation from linearity up to the used flux (10$^5$ n/cm$^2$.s).

\subsection{Position resolution}
\begin{figure}[H]
\centering
\includegraphics[width=90mm]{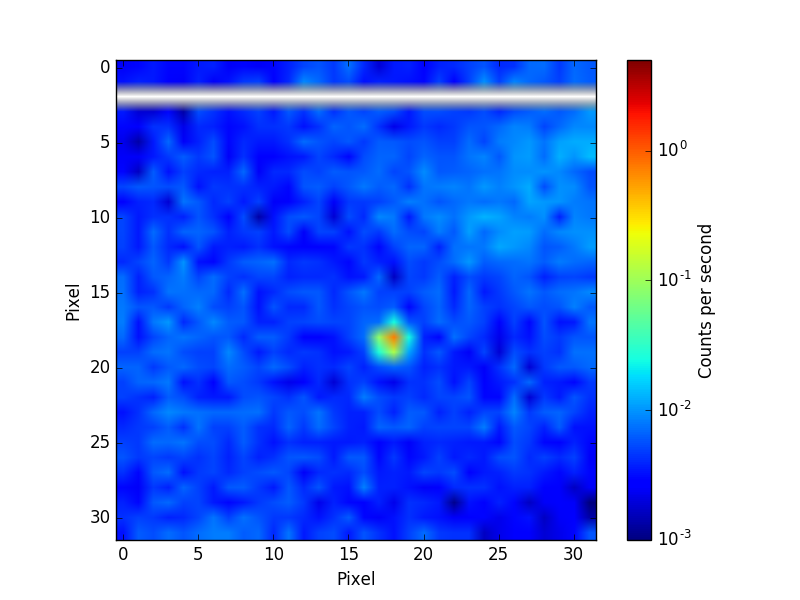}
\caption{Two-dimensional image measured by GEM monitor at R2D2 at IFE using a neutron beam size of 1x1mm$^2$ at 2.4\AA. The white strip is a dead electronics channel. One pixel size is 3.12 mm.
\label{CDT_image}}
\end{figure}

\begin{figure}[H]
\centering
\includegraphics[width=90mm]{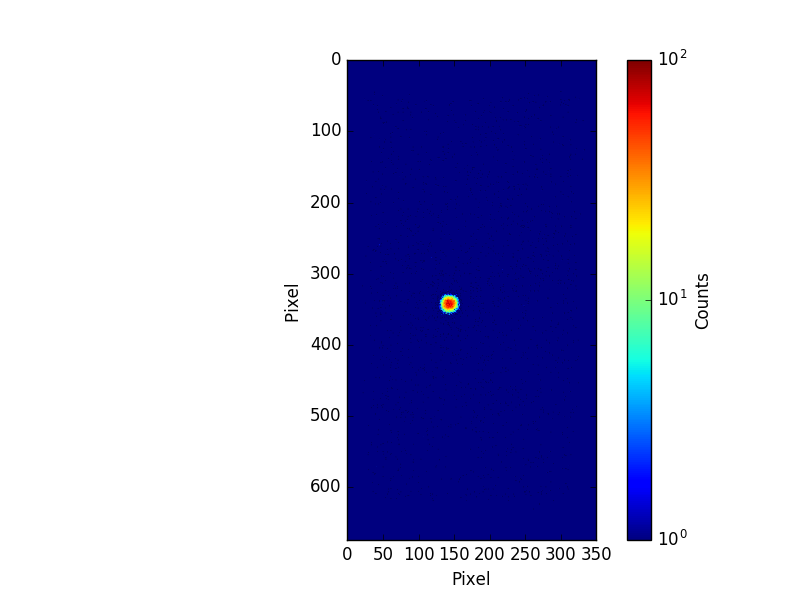}
\caption{Two-dimensional image measured by MIR2D at R2D2 at IFE using a neutron beam size of 1x1mm$^2$ at 2.4 \AA. The pitch size is 1mm x 1mm.
\label{MIR-2D_image}}
\end{figure}

Finally, one important characteristics of a position sensitive detector is its spatial resolution defined
as FWHM (full width at half maximum). For such characterization the beam should be as narrow as possible. This was achieved using a collimated beam size with two set of slits. This makes sure that the beam is not widened due to divergence.  The measuring time was  1 hour for the GEM monitor and for 18 hours for the Mirrotron-2D.  The estimated FWHM value of the GEM is  3.1 mm and 1.3 mm for Mirrotron-2D  . The 2D-image measured by GEM and Mirrotron-2D are  shown in figure~\ref{CDT_image} and~\ref{MIR-2D_image}  respectively.

\section{Conclusions}

 Different types of neutron beam monitors (MWPC, GEM, scintillators and fission chamber) purchased from different companies (ORDELA, Mirrotron, Scintillator from Quantum Detectors, LND and CDT) have been tested using a neutron source and different gamma sources and using the neutron R2D2 beamline at IFE-Norway. The efficiency, attenuation and gamma sensitivity have been studied. It is striking the high level of attenuation shown by all but one of the monitors. This is related to the used materials in each monitor (mainly the Al window or kapton in case of the used position sensitive monitors ). For instance this factor could be reduced by reducing the window thickness. 
The fission chamber from LND showed the lowest gamma sensitivity thanks to the large Q value ($\approx$200 MeV) of the neutron-induced fission reaction. At the moment, the monitor efficiency is far lower than the beam attenuation.
This work provides useful information for users at different facilities in order to choose a proper monitor which best satisfies their requirements. Based on these measurements some of these beam monitors will be customized in order to match the ESS instrument requirements. Future developments on monitors should focus on reducing the level of scattering and attenuation of the beam. 

\section{Acknowledgments}
This work is funded by EU Horizon2020 framework, BrightnESS project 676548 and was carried out in part at the Source Testing Facility, Lund University. We would like to thank K.Berry and N.Rhodes for the useful discussions; the manufacturers for their openness.


\medskip
\bibliographystyle{unsrt}
\bibliography{BM_bibo}
\end{document}